\documentclass[12pt]{article}
\usepackage{amsmath}
\usepackage{amssymb}

\title{LHC card games: bringing about retrocausality?}

\author{ Z.~K.~Silagadze \\
Budker Institute of Nuclear Physics and \\
Novosibirsk State University, 630 090, Novosibirsk, Russia }
\date{}

\begin{document}

\maketitle

\begin{abstract}
The model of Nielsen and Ninomiya claims that ``the SSC (Superconducting
Supercollider) were stopped by the US Congress due to the backward causation 
from the big amounts of Higgs particles, which it would have produced, if it 
had been allowed to run''. They also proposed to play a card game and if the
``close LHC'' card is drawn (with probability $\sim 10^{-6}$), really close 
LHC on the eve of Higgs particle discovery to avoid more severe bad luck.
Crazy? Probably. But paraphrasing Salvador Dali, if you believe that you
and me are smarter in physics than Nielsen and Ninomiya, don't read this 
article, just go right on in your blissful idiocy. Therefore, I will try
to make sense of backward causation. It turns out that not only the 
backward causation makes perfect sense in some models of possible reality,
but that Nielsen and Ninomiya really have a chance to close LHC by a card 
game. The only thing they need is to be smart enough to manage to develop 
their theory up to the level of brilliance beginning from which it becomes 
a part of the fabric of reality. We hope, however, that they will use their 
outstanding abilities to bring about some more interesting future.  

\end{abstract}

\section{Introduction}
In a number of seemingly crackpot papers \cite{1,2,3,4,5}, Nielsen and
Ninomiya have developed a theory that is undoubtedly crazy. But craziness is 
not always the reason to reject a theory. If it were we would not have 
neither relativity nor quantum mechanics today. Nielsen and Ninomiya base
their theory over Feynman's approach to quantum mechanics which at first
sight also seems crazy. Freeman Dyson recollects: {\it Thirty-one years ago, 
Dick Feynman told me about his ``sum over histories'' version of quantum 
mechanics. ``The electron does anything it likes,'' he said. ``It just goes 
in any direction at any speed, forward or backward in time, however it 
likes, and then you add up the amplitudes and it gives you the 
wave-function.'' I said to him, ``You're crazy.'' But he wasn't} \cite{6}.
Therefore, let us take closer look to the idea of Nielsen and Ninomiya.

According to Feynman's approach to quantum mechanics \cite{7}, to find the 
probability $P_{ba}$ for a quantum system's transition from a state $|a>$ 
to a state  $|b>$, Nature applies three simple rules:
\begin{itemize}
\item  Explore all ``paths'' connecting $|a>$ to $|b>$.
\item A complex number of unit magnitude
\begin{equation}
{\cal{A}}(\mathrm{Path})=\exp{\left (\frac{i}{\hbar}S(\mathrm{Path})\right )},
\label{eq1}
\end{equation}
called the amplitude, is prescribed to each path. 
\item The probability $P_{ba}=|{\cal{A}}|^2$ is proportional to the squared 
modulus of the complex number ${\cal{A}}$ which is just the sum
\begin{equation}
{\cal{A}}=\sum\limits_{\mathrm{all paths}} {\cal{A}}(\mathrm{Path}).
\label{eq2}
\end{equation}
\end{itemize}

The quantity $S(\mathrm{Path})$ plays the central role in the above scheme
and is called action. Usually it is assumed that the action is a real number.
Therefore all paths connecting $|a>$ to $|b>$ are equivalent in a sense that 
all these complex numbers have unit magnitude and hence if we restrict the 
system in such a way that the only one path connecting $|a>$ and $|b>$ remains
then $P_{ba}(\mathrm{Path})=\mathrm{const.}$ irrespective of the path chosen. 
However, the phases of amplitudes may be different and when there are many 
paths connecting $|a>$ and $|b>$, the probability $P_{ba}$ depends strongly
on how these amplitudes interfere with each other. 

The novelty of Nielsen and Ninomiya's approach is that they ask what will
happen if we allow imaginary part in the action. Then the equality of paths
is broken down because $P_{ba}(\mathrm{Path})\sim \exp{[-2\,Im(S)/\hbar]}$
and certain trajectories whose actions have large positive imaginary parts 
will be highly suppressed.

There is nothing bizarre or unprecedented in considering imaginary parts
of the actions. It was shown by Schwinger long ago \cite{8} that the 
effective action for a constant electric field develops a positive imaginary
part. Schwinger interpreted this imaginary part as an indication that the QED 
vacuum in a background electric field is unstable and in strong enough fields
a kind of vacuum electrical breakdown due to spontaneous electron-positron  
pair production takes place \cite{9}.

The imaginary part of the action arises also quite naturally in the WKB 
description of quantum tunneling. In a tunneling event there are two separated
classical turning points which are joined by a classically forbidden 
trajectory.
The probability of tunneling is related to the imaginary part of the action 
for this classically forbidden trajectory. The most impressive application
of this kind of the imaginary part of the action is, probably, the derivation 
of the Hawking radiation as a tunneling event \cite{10}.

Of course, it will be very interesting to experimentally confirm Schwin\-ger's 
insight about the imaginary part of the Euler-Heisenberg-Schwinger (EHS) 
action because there are some theoretically embarrassing moments about this
effective action. Namely,  the effective EHS action exhibits mysterious 
statistics reversal: beginning from a microscopic theory of fermion pair 
vacuum fluctuations we end with the effective EHS action which have a form 
typical for bosons in a thermal bath, while for the spin-0 effective action
the form is typical for fermions in a thermal bath \cite{11}. Besides, the
temperature of the thermal bath itself differs from the corresponding 
Hawking-Unruh temperature by a factor two and nobody knows why \cite{11}. 
Interestingly, Schwinger mechanism seems to be testable experimentally in
graphene which mimics effective relativity with massless fermions \cite{12}.

The use of imaginary part of the action by Nielsen and Ninomiya is, however,
conceptually different. They want this imaginary part to suppress all 
histories of the universe except a few, ideally just one history, with 
preselected initial conditions. Such a super-theory will be 
superdeterministic: if you know exact form of action, both real and imaginary 
parts, you can predict everything which will happen or happened in this 
universe.

Of course Nielsen and Ninomiya do not yet have such a theory and I doubt they 
ever will. However, they develop an embryonic version of it which predicts 
a large positive imaginary part of the action for every history in which 
large numbers of Higgs bosons are produced. Therefore, such histories are
exponentially suppressed and never realized.

Up to this point, Nielsen and Ninomiya's theory, although extravagant, is  
completely scientific. Moreover, it has a great virtue which every good 
scientific theory is supposed should have: it is falsifiable. It makes a 
strong prediction that LHC will never succeed in production of large numbers 
of Higgs bosons.

However, Nielsen and Ninomiya go beyond this point. They claim that the real
reason why the SSC (Superconducting Supercollider) was canceled by USA 
Congress was that if it had been allowed to run it would have produced big
amounts of Higgs particles and such a history is suppressed by the 
corresponding large imaginary part in the action. Similarly, a bad luck is
awaiting to LHC too. To avoid severe potentially harmful accidents which can 
``naturally'' stop LHC if their theory is true, Nielsen and Ninomiya suggest 
to play a card game. Take about one million cards most of which say ``Go on
with LHC, discover Higgs boson and be happy'' but few of them prescribe some
restrictions on allowed LHC energy or luminosity. And just one card says 
``Stop LHC and never turn it on''. If the ``Close LHC'' card is drawn, this 
improbable event, according to Nielsen and Ninomiya, will indicate that their
theory of imaginary action is true and we must really close LHC, otherwise
some natural or political catastrophic event will do it instead of us.

What we can say about this borderline-crackpot suggestion? First, it does not
follow from the Nielsen and Ninomiya's particular form of the imaginary 
action. This action simply says that any history with large amounts of produced
Higgs particles is extremely improbable. But there is nothing in it which
indicates that the history in which LHC is closed by Nielsen and Ninomiya's
card game is more probable than the history in which LHC is simply blown up
by terrorists: in both case the LHC will be unable to produce Higgs particles
and frankly speaking the second case gives even more guarantees that it never 
will. Therefore, Nielsen and Ninomiya here are implicitly assuming
that we have free will to choose between histories whose imaginary actions
are close enough and somehow the histories with less harm to humans are more 
probable.    

Besides, there is one logical loophole in the argumentation of Nielsen and 
Ninomiya. Suppose their theory is true and all universe histories with big
chunks of produced Higgs particles are exponentially suppressed. Does then
it mean that LHC will never be able to produce significant amount of Higgs 
particles? Not necessarily, in light of Multiverse theory.

The Multiverse theory \cite{14,15,16} is in every bit more miraculous 
than Nielsen and Ninomiya's theory. For example it predicts that about
$10^{10^{29}}~\mathrm{m}$ away there is an exact copy of your \cite{15}
reading the copy of this article. Nevertheless, the Multiverse theory is 
considered as completely respectable, even mainstream theory and many
prominent physicists are confident in it. For example, Martin Rees is
sufficiently confident about the multiverse to bet his dog's life on
it, Andrei Linde is ready to bet his own life, and Steven Weinberg has
just enough confidence to bet the lives of both Andrei Linde and Martin 
Rees's dog \cite{14}.

The Multiverse theory has two implications for the Nielsen and Nino\-miya's
hypothesis. Firstly, it asserts that there is a parallel universe in which
the Nielsen and Ninomiya's card game has been already played and as a result
of the game's outcome the analog of LHC was closed (however this does not 
prove that Nielsen and Ninomiya's imaginary action theory is true. Even 
without this theory, there is a non-zero probability that certain authorities 
can make a foolish decision). Secondly, even if Nielsen and Ninomiya's theory
is true, there exists a universe in which the exponentially small probability
that LHC will be successful is realized. To the delight of high-energy 
physicists, there is no reason why this universe could not be our own.

Therefore, I think, we have every reason not to worry about LHC card games.
The Multiverse theory kills the super-determinism of the Nielsen and 
Ninomiya's proposal. In fact, we could not compute our own future even if
we had the knowledge of the entire state of multiverse, because there are 
infinitely many copies of us and our universe and their histories will
eventually deviate, but there is no way to determine what particular copy
you and me belong to \cite{15}.  

However, there is one aspect of the Nielsen and  Ninomiya's theory which 
deserves to be further scrutinized. They interpret a possible stoppage of
LHC by a card game as an example of backward causality, that some event in 
future (production of Higgs particles) prearranges conditions today in order
not to happen. Does this make sense? We need some analysis of the notion of 
causality to answer this question. 

\section{What is time?}
Causality is intimately related with the notion of time. Therefore, the first
question we should try to answer is about the origin of time. Most of the 
modern physicists are, perhaps, pretty sure that they know what is time: it 
is just forth dimension of space-time different from the spatial dimensions in
a subtle way summarized in the pseudo-euclidean character of the metric.
This attitude goes back to Minkowski: ``Henceforth space by itself, and 
time by itself are doomed to fade away into mere shadows, and only a kind of 
union of the two will preserve an independent reality'' \cite{17}.

Ordinary man, however, will not object if I say that space and time are two 
big differences. ``What is meant by the word 'Time'? There is no scientific 
answer to this question. What is meant by the word 'Space'? Here, rational 
thought may possibly provide us with an answer. Yet a connection exists 
between Destiny and Time, and also between Space and Causality. What, then, is 
the relationship between Destiny and Cause? The answer to this is fundamental 
to the concept of depth experience, but it lies beyond all manners of 
scientific experience and communication. The fact of depth experience is as 
indisputable as it is inexplicable'' \cite{18}. As we see, Spengler thinks 
that it is particularly difficult to understanding the concept of time. How 
fortunate that you and me can live, and sometimes even live very successfully, 
without thinking on such complicated issues. ``The active person lives in the 
world of phenomena and with it. He does not require logical proofs, indeed he 
often cannot understand them'' \cite{18}. 

However, the concept of background Minkowski space-time was extraordinary
successful in the realm of physics. How much has it enlightened the enigma
of time? The most complete exposition of what the special theory of relativity
has to say about space-time can be found in the great treatise of Alfred Robb
``A Theory of Time and Space'' \cite{19}. He used an axiomatic approach and,
in fact, has been nicknamed ``the Euclid of Relativity'' \cite{20}. The basic
concept, introduced by Robb, which is at the heart of causal structure of
special relativity, is the notion of ``Conical Order'' \cite{20}. This notion
emphasizes an important difference between simultaneity and succession of
spatially separated events. Einstein bewildered contemporaries by showing
that simultaneity is relative. Conical order, however, enables to introduce
an absolute succession between events and define the notions of {\it after} 
and {\it before} which are not relative but absolute. Physicists were so
mesmerized by Einstein's great discovery that Robb remains up to now an
forgotten hero of relativity. Nevertheless, as the founder of causal theory
of time \cite{21}, the contribution of Alfred Robb in relativity deserves to
be considered as important as achievements of Einstein and Minkowski.

As for Einstein, the devil seduced him to develop general theory of
relativity to obscure again the problem of time. 

Space-time in general relativity is a four-dimensional pseudo-Riemannian 
manifold. Its symmetric metric tensor $g_{\mu\nu}$ is dynamical in the sense 
that it is determined by matter distribution according to the Einstein 
equations
 \begin{equation}
R_{\mu\nu}-\frac{1}{2}g_{\mu\nu}R=\frac{8\pi G}{c^4}T_{\mu\nu}.
\label{eq3}
\end{equation}
If the energy-momentum tensor $T_{\mu\nu}$ in the right-hand side of this 
equation is related to the motion of a point particle then equation of motion
of this particle follows from (\ref{eq3}) itself due to Bianchi identities.
This equation simply says that the particle moves along a geodesic. 
Therefore, beautifully and miraculously, gravity in general relativity emerges 
as a manifestation of space-time curvature. Usually this aspect of general 
relativity is considered as revolutionary. However, it is not. As pioneered 
by Cartan and Friedrichs, Newtonian gravity also can be cast in a generally 
covariant form in which both characteristic features of general relativity, 
gravity as spacetime curvature and dynamical metric, are realized \cite{22}. 

What is really specific to general relativity is that the geometry of tangent 
space to any point is Minkowskian, while in Newton-Cartan theory tangent
spaces have Galilean geometry. This makes a big difference as far as the 
issue of time is concerned.

Newton-Cartan space-times are globally hyperbolic with absolute time and
fixed causal structure. In the case of general relativity, the nice 
Minkow\-ski geometry  of tangent spaces allows to introduce conical order and 
hence causal structures in these tangent spaces. However, it is not clear 
whether for every pseudo-Riemannian space-time, which follows from the 
Einstein's equations (\ref{eq3}), these local causal structures could be 
integrated in a global definition of time.

Indeed, G\"{o}del found a solution that has closed timelike curves through 
every event \cite{23}. Therefore, our intuitive understanding what time is 
breaks down in the G\"{o}del's universe as we cannot define the meanings of 
'before' and 'after' for events globally.

Quantum mechanics brings another flavor in the problem of time. Time plays 
a special role in quantum theory. Unlike spatial coordinates, time can not be 
represented by a self-adjoint operator and, therefore, is not a physical 
observable in the normal sense. Pauli's argument that this is indeed so goes
as follows \cite{24}. Suppose there exist a self-adjoint operator $\hat T$
canonical conjugate to the Hamiltonian $\hat H$. Then we should have
\begin{equation}
[\hat H,\,\hat T]=i\hbar.
\label{eq4}
\end{equation}
But this commutation relation shows that if $|E>$ is an energy eigenstate, 
then 
$$\exp{\left (\frac{i}{h}E_1\hat T\right )}|E>$$
is also energy eigenstate but with energy eigenvalue $E-E1$. Therefore, any
such operator necessarily implies that the spectrum of Hamiltonian is not 
bounded from below which excludes most physically interesting systems since
they are assumed to have a stable ground state.

Although Pauli's ``theorem'' is not mathematically rigorous and there is a
loophole in it \cite{24,25}, time undoubtedly plays a subtle role in quantum 
mechanics. In general relativity, space-time coordinates are mere labels
attached to events and true physical results are assumed to be independent of 
choices of such coordinates. This fact creates problems for interpretation of
the meaning of time already at classical level and it is not surprising that 
the problem of time only becomes more acute when we try to merge quantum 
mechanics and general relativity in quantum gravity \cite{26}.

To conclude, ``Although time is a concept that attracted and occupied the 
thoughts of a countless number of thinkers and scholars over centuries, its 
true nature still remains wrapped in a shroud of mystery'' \cite{27}.

\section{Are our theories casual?}
Anybody who thinks that causality and determinism are simple and trans\-parent 
notions should consult John Earman's {\it A primer on determinism} to find
out that determinism is a vague concept and causation is a truly obscure one
\cite{28}. We will not try to enter any deeply into these wilds but only 
scratch the surface of the problem.

It is usually thought that the classical physics is a paradise for 
determinism and spacial relativity and quantum physics have spoiled this 
harmony. Just the contrary. Newtonian world is quite hostile to 
determinism, quantum mechanics is more deterministic than classical 
mechanics, and special relativity is our best theory where the most dreams
of determinism can be realized. To these dreams, however, general relativity 
poses new grave challenges \cite{28}. 

let us briefly indicate just a few examples of acausal behavior of seemingly 
benign Newtonian systems. 

Imagine a system of two equal masses $M$ moving in the $x-y$ plane under 
Newton's inverse square force law and the mirror replica of this binary placed
symmetrically at a large distance. The fifth mass $m\ll M$ is placed on the 
$z$-axis which goes through the centers of mass of the planar binary systems.

In his efforts to solve the century-old problem of noncollisional 
singularities, Xia was able to show \cite{29} that there exists a set of the 
initial conditions for which the four bodies, constituting binaries in the 
above construction, will escape to spatial infinity in a finite time, while 
the fifth small body will oscillate back and forth between these binaries 
with ever increasing speed.

However classical mechanics is time reversal invariant and the time reverse 
of the Xia's construction is an example of ``space invaders'' \cite{30}, 
particles appearing from spatial infinity  in a surprise attack without any 
apparent cause. 

Another example is given by P\'{e}rez Laraudogoitia's beautiful supertask
\cite{31}. An infinite set of identical particles is arranged in a straight 
line. The distance between  the particles and their sizes decrease so that 
the whole system occupies an interval of unit length. Some other particle of 
the same mass approaches the system with unit velocity. A wave of elastic
collisions goes through the system in unit time and all particles come to 
rest after this collision supertask is over. The time reversal of the
P\'{e}rez Laraudogoitia's supertask implies indeterminism because it is the 
following process: a spontaneous self-excitation propagates through the 
infinite system of balls at rest causing the first ball to be ejected with 
some nonzero velocity.

Surprisingly, quantum mechanics is more friendly to determinism than the 
Newtonian mechanics \cite{32}. Supertasks are possible in quantum mechanics 
too \cite{33}, and there are quantum supertasks in which the spontaneous 
self-excitation of the ground state is allowed \cite{34}. However, 
pathologies disappear and the quantum mechanical supertasks  are better 
behaved than their classical counterparts if one demands normalizability of 
the state vector \cite{34}.

Let us mention also Norton's fascinating example \cite{35,36}. Suppose an 
equation of motion for a Newtonian particle is (Norton provides a concrete
dome-like construction which leads to this equation of motion)
\begin{equation}
\frac{d^2r}{dt^2}=k\sqrt{r},
\label{eq5}
\end{equation}
where $k$ is some dimensionful constant. If the particle is initially at rest 
at the origin $r=0$, then the obvious solution of (\ref{eq5}) is $r(t)=0$. 
But, surprisingly, there is also a whole class of other solutions
\begin{equation}  
r(t)=\left \{ \begin{array}{c} 0,\;\;\mathrm{if} \;\; t\le T, \\ \\
\frac{k^2}{144}(t-T)^4, \;\;\mathrm{if} \;\; t> T. \end{array} \right . 
\label{eq6}
\end{equation}
It is easy to check that (\ref{eq6}) is a solution of (\ref{eq6}) with 
required initial conditions for any $T>0$. But then we have an amusing 
situation: a particle sitting at the apex of the Norton's dome in the 
gravitational field begins to move spontaneously, without any cause, at an 
arbitrary time $t=T$, in some arbitrary radial direction.

Norton's dome illustrates well that our implicit beliefs in causality demands 
some kind of smooth structures (Norton's dome is $C^1$ but not $C^2$ at the 
apex). Non-smooth structures open a Pandora box of very strange objects like 
the Devil's staircase  \cite{37}. Devil's staircase implies a possibility for 
a particle to advance forward with a continuous constant velocity which is 
zero nearly everywhere! Nevertheless, such Cantor functions are not merely
pathological oddities as they appear naturally in various areas of mathematics 
and mathematical physics. 

The problem with smooth structures is that for 4-dimensional manifolds there
exist different, not diffeomorphic to each other, smoothnesses and what is 
smooth in one smoothness is not smooth in another. However, there is no 
physical ground to favor one smoothness over another \cite{38,39}.

To conclude, causality is not a simple notion. Over the centuries there were 
hard efforts to distillate and make transparent the principle of causality 
and there is ``such a history of persistent failure that only the rashest 
could possibly expect a viable, factual principle still to emerge'' \cite{40}.

\section{Backward causality}
At first sight backward causality does not makes sense. However, let us take
the following example \cite{41}. Suppose your son was on a ship that has 
gone down in the ocean two hours previously according to the radio broadcast
you just listened. The broadcast have mentioned that there were a few 
survivors. I expect you immediately to utter a prayer to almighty God that
you son should have been among the survivors, and I affirm that such 
a behavior is the most natural thing in the world.  

However, it is logically impossible to alter the past and such a prayer is
blasphemous according to orthodox Jewish theologians \cite{41}, because we
are asking God to perform an impossible thing.

Christian tradition, on the other hand, will bless such a prayer. Are 
Christians less logical and at error in this case?

If you don't like the theological example, Dummett provides a magical one
\cite{41}. Imagine a tribe that has the following initiation ritual. Every
second year the young men have to go off for six days. They travel for
the first two days to some isolated place. Then they hunt lions to confirm
their manhood. Last two days they spent on the return journey.

The chief of the tribe believes he can influence the outcome of the test if
he dance and all the time he eagerly performs this ritual. The weird thing 
is that the chief continues these dances for the whole six days. In our 
opinion, the dancing can not bring about the young men's bravery and, 
therefore, the chief has wholly mistaken system of causal beliefs. In 
particular, we consider as especially absurd an idea to continue the dance
for last two days after the lion hunting is already over. Can we persuade 
the chief on the empirical ground that his behavior is absurd? Dummett argues
that we can not. The interpretation of empirical data depends on some deeply 
rooted conceptual beliefs or prejudices. ``If we were as convinced as he is 
of the existence of sorcerers and of mysterious powers, instead of believing 
in  so-called natural causes, his inferences would seem to us perfectly 
reasonable. As a matter of fact, primitive man is no more logical or 
illogical than we are. His presuppositions are not the same as ours, and  
that is what distinguishes him from us" \cite{42}.  

One of the presuppositions which we take for granted is that the past is
fixed in every detail and can not be changed. I agree that under such 
presupposition backward causality does not make sense. However, is it 
an absolute logical necessity to stick to this prejudice?

\section{A model for backward causality}
In fact, such a model was suggested in \cite{43} and is based on a wild but
not impossible (especially in the multiverse theory) idea that we live in a 
computer simulation \cite{44,45}. During all the history, ancestor worship 
was very strong religious tradition. it is not unbelievably unrealistic that 
enormous amounts of computing power will be available in the future. ``One 
thing that later generations might do with their super-powerful computers is 
run detailed simulations of their forebears or of people like their forebears. 
Because their computers would be so powerful, they could run a great many such
simulations. Suppose that these simulated people are conscious (as they would
be if the simulations were sufficiently fine-grained and if a certain quite
widely accepted position in the philosophy of mind is correct). Then it could
be the case that the vast majority of minds like ours do not belong to the
original race but rather to people simulated by the advanced descendants of
an original race'' \cite{44}.

Now suppose that this high tech substitute of the ancestor worship is 
self-adaptive. I mean that the rules of this game (which we call natural 
laws) are not fixed forever but can change defending the participants' 
creative  output. So to say, we are co-creators of this world not just 
passive actors. Of cause such a world view is strongly anti-Copernicean,
contrary to the last centuries scientific mainstream, but I find nothing
particularly impossible in it.

In such a virtual reality the past is not fixed in every detail, otherwise
it would be a foolish waste of computer memory. Backward causality is a 
natural thing in such a universe: some details of the past are fixed only 
when we pay our attention to them from the future.

\section{Concluding remarks}
Can Nielsen and Ninomiya bring about the closure of LHC by a card game?
Yes they can if we live in a kind of virtual universe outlined above. 
However, our experience with scientific exploration of the world indicates
that there is some highly aesthetic underling principles of the fabric of this 
computer game. Therefore, Nielsen and Ninomiya have to work hard to meet this
standards. But are such efforts worthwhile if only LHC closer is at the 
stake?

\end{document}